\documentclass[11pt]{article}

\usepackage[a4paper,margin=1in]{geometry}
\usepackage{lmodern}
\usepackage[T1]{fontenc}
\usepackage[utf8]{inputenc}
\usepackage{amsmath,amssymb}
\usepackage{graphicx}

\usepackage{booktabs}
\usepackage{siunitx}
\usepackage[hidelinks]{hyperref}
\usepackage[numbers,sort&compress]{natbib}
\usepackage{caption}

\usepackage{indentfirst}

\newcommand{\Bud}{\mathcal{B}}
\newcommand{\Budad}{\Bud_{\mathrm{ad}}}
\newcommand{\Buddc}{\Bud_{\mathrm{dc}}}
\newcommand{\TL}{\mathrm{TL}}
\newcommand{\Cspeech}{\mathcal{C}}

\title{\bfseries
Causal limits and optimal allocation for passive frequency-selective hearing protection
}

\author{%
\parbox{\textwidth}{%
\centering
Geonhwi Hwang$^{1,*}$,
Jaewoo Lee$^{2}$,
Hyunjun Kim$^{3}$,
and Hyeonwoo Na$^{4}$\\[6pt]
$^{1}$ Department of Applied Physics, Hokkaido University, Sapporo 060-8628, Japan\\
$^{2}$ CHA University School of Medicine, Gyeonggi-do 13488, Republic of Korea\\
$^{3}$ School of Computing, Korea Advanced Institute of Science and Technology, Daejeon 34141, Republic of Korea\\
$^{4}$ Division of Mechanical and Space Engineering, Hokkaido University, Sapporo 060-8628, Japan\\[3pt]
$^{*}$ Corresponding author: theprex2@eis.hokudai.ac.jp
}%
}

\date{}

\begin{document}

\maketitle
\thispagestyle{empty}

\begin{abstract}
\noindent
\textbf{Introduction:} Passive Helmholtz-resonator earplugs aim to
attenuate hazardous noise while keeping speech audible; the air volume of
an ear-sized shell is finite, and we treat it as the resource a design
spends.
\textbf{Methods:} For a vented bore transparent at zero frequency and
loaded by reactive side branches, causality fixes a conserved integral of
transmission loss. We derive it in device variables, add viscous and
flanking constraints, and allocate the budget by hazard-weighted
water-filling with a continuous speech-cost price.
\textbf{Results:} The audio-band budget $\Budad=(20/\ln10)\pi^{2}V/2cS$ is
set by the ratio of total cavity volume $V$ to bore area $S$; occluding
plugs are exempt. The $f^{2}$ price signal favours frequencies well above
the source peak. Replacing strict transparency by a speech-cost price
gives an exchange rate of about \SI{1.3}{\decibel} of protection per
decibel of speech cost under the present model, with the budget setting
where it saturates.
\textbf{Discussion:} Across three form factors, low-frequency-selective
designs stay near a decibel while hazard-matched mid/high-frequency
designs reach twelve. The binding variables are bore area and speech
allowance, not resonator count.
\end{abstract}

\vspace{2pt}\noindent\textbf{Keywords:} hearing protection; Helmholtz
resonator; causality sum rule; acoustic metamaterials; impulse noise

\section{Introduction}

Hearing protection in hazardous-noise environments carries a
long-standing dual requirement: attenuate the threat, but keep speech and
warning signals audible. Weapon noise is a principal such threat, reaching
peak levels near \SI{160}{\decibel} at the shooter's
ear~\cite{wicher2019gunshot,flamme2009estimates} and defining the impulse
damage-risk criteria
that protectors are designed against~\cite{chan2001evaluation}. Electronic
level-dependent protectors address the dual requirement at the cost of
batteries, and a growing body of work therefore asks
whether a purely passive, frequency-selective structure can do the same.
Helmholtz-resonator (HR) ``meta-earplugs'' are the leading candidate.
Critically coupled HR arrays have been shown to suppress the occlusion
effect~\cite{carillo2022,carillo2023}, the resulting comfort gains have
been evaluated on human listeners~\cite{carillo2025}, and tuning the
reflection coefficient at the earplug's medial surface has recently been
shown to improve low-frequency noise reduction~\cite{carillo2026}. These
devices sit within the broader acoustic-metamaterial programme, in which
locally resonant elements realise effective material parameters
unavailable in natural
media~\cite{liu2000locally,fang2006ultrasonic,ma2016acoustic,cummer2016controlling};
related side-branch, perfect-absorber and panel concepts recur throughout
the transmission literature~\cite{huang2019embedded,jimenez2017,romero2016}.

Separately, a rigorous theory of what passive structures cannot do
has matured. Causality and passivity impose sum rules relating a
structure's broadband scattering response to its static compliance and
inertia: Rozanov's thickness--bandwidth bound for radar
absorbers~\cite{rozanov2000}, extended beyond the Rozanov criterion for
thin absorbers~\cite{firestein2023sumrule}, its acoustic counterparts for
absorbing layers~\cite{yang2017,acher2009fundamental}, integral identities
for one-dimensional scattering coefficients~\cite{norris2018},
Herglotz-function sum rules
that predict the minimum space required for a target absorption or
transmission-loss spectrum~\cite{meng2022}, and most recently an acoustic
analogue of the Baldin sum rule locking the extinction integral to static
effective mass and stiffness~\cite{qu2026}, since extended to two-port
monopole--dipole scattering through a duality symmetry~\cite{qu2025generalized}.
General constraints on passive systems of this kind are reviewed in
Ref.~\cite{bernland2011}.

These two literatures have not met. The bounds are stated in general
scattering variables; the devices are specified in cavity volumes, neck
diameters and bore diameters. This paper closes that gap for the
ear-protector geometry and asks the question a device designer faces:
given an ear-sized air volume, a specified threat, and a requirement that
speech remain usable, what attenuation spectra are physically available,
and which of them are worth building?

The conserved integral itself is not new: Equation~(\ref{eq:budget})
specialises the scattering identities of Refs.~\cite{norris2018,meng2022}
to the bore-plus-side-branch geometry, written in device variables, and
the contemporaneous general relation of Ref.~\cite{qu2026} shares its
$\omega^{-2}$ kernel and causal origin. What the integral implies for
device design is the substance of this paper: (i) an explicit statement of
which protectors the budget binds, which removes an apparent conflict with
measured low-frequency gains in occluding meta-earplugs; (ii) a third,
non-causal constraint, a flanking path, that closes the design space from
above and that the scattering bounds do not see; (iii) a source-matched,
hazard-weighted allocation rule; (iv) the replacement of the
all-or-nothing transparency constraint by a continuous speech-cost price,
which converts a prohibition into an exchange rate; and (v) a design atlas
that locates five commercially motivated archetypes inside the resulting
space.

Throughout we deliberately avoid tying the analysis to a particular
shell. A device is specified only by its total cavity volume $V$,
the number $N$ of cavities it is divided into, and the open bore area
$S$; no cross-section, sector count or moulding strategy is assumed.
Three representative classes are carried through the paper
(Table~\ref{tab:forms}): a canal plug, a deep canal plug and a
concha-plus-canal mould. They span roughly an order of magnitude in
budget and bracket what is realistic for an in-ear device.

\section{The volume budget}
\label{sec:budget}

\subsection{A conserved integral in device variables}

Consider a rigid bore of cross-sectional area $S$ carrying sound to the
eardrum, loaded at a sub-wavelength junction by $N$ side branches of
total acoustic admittance $Y_{b}(\omega)=\sum_{i}Z_{i}^{-1}(\omega)$,
where $Z_{i}=R_{i}-i\omega M_{i}+i/(\omega C_{i})$ is the lumped
impedance of resonator $i$ (convention $e^{-i\omega t}$), the standard
Helmholtz-resonator description~\cite{kinsler2000fundamentals,kim2006hrarray}.
The acoustic compliance of cavity $i$ is $C_{i}=V_{i}/\rho c^{2}$ and its
inertance is $M_{i}=\rho L_{\mathrm{eff},i}/A_{i}$. The pressure
transmission coefficient is
\begin{equation}
  t(\omega)=\Bigl[1+\frac{1}{2}Z_{0}Y_{b}(\omega)\Bigr]^{-1},
  \qquad Z_{0}=\rho c/S .
  \label{eq:t}
\end{equation}

Three properties follow. Passivity ($\operatorname{Re}Z_{i}\ge 0$) gives
$|t|\le 1$. Causality makes $t$ analytic in the upper half
$\omega$-plane, and the zeros of $t$, the roots of $Z_{i}=0$, lie at
$\operatorname{Im}\omega=-R_{i}/2M_{i}<0$ for any dissipative branch, so
$h(\omega)\equiv-\ln t(\omega)$ is analytic there too. Finally $t\to1$ at
both ends of the spectrum: at $\omega\to0$ each branch is blocked by its
compliance, at $\omega\to\infty$ by its inertance. Expanding about the
origin,
\begin{equation}
  h(\omega)=-\frac{i}{2}\,\omega Z_{0}C(0)+\mathcal{O}(\omega^{2}),
  \qquad C(0)=\sum_{i}C_{i}(0),
  \label{eq:lowfreq}
\end{equation}
with $V=\sum_{i}V_{i}$ the total cavity volume. Only the total
appears: the low-frequency behaviour is blind to how the volume is
partitioned among cavities.

\begin{figure}[t]
  \centering
  \includegraphics[width=\linewidth]{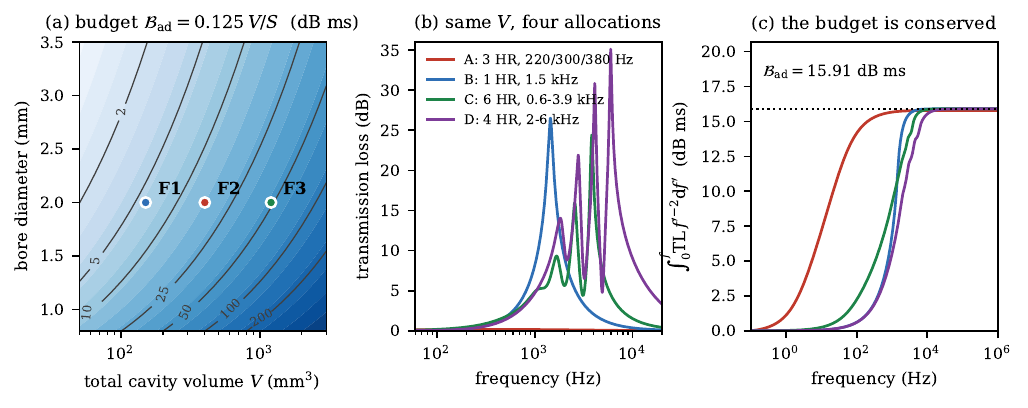}
  \caption{(a) The audio-band budget $\Budad$ of Eq.~(\ref{eq:rule}) over
  the design plane, with the three representative classes of
  Table~\ref{tab:forms} marked; contours are labelled in
  \si{\decibel\,\milli\second}. Because $\Budad\propto S^{-1}\propto
  d^{-2}$, reducing the bore diameter raises the budget as effectively as
  enlarging the volume, at no cost in packaging space.
  (b) Transmission loss of four dissimilar allocations of the same
  \SI{400}{\cubic\milli\metre}. (c) Their running integrals
  $\int_{0}^{f}\TL f'^{-2}\mathrm{d}f'$, all converging to $\Budad$
  (dotted). The \SIrange{220}{380}{\hertz} array spends its budget below
  \SI{100}{\hertz}, in an overdamped and inaudible tail.}
  \label{fig:fig1}
\end{figure}

The compliance that anchors the expansion is the static one.
Static compression of the gas in a thermally conducting cavity is
isothermal rather than adiabatic, so $C_{i}(0)=\gamma V_{i}/\rho c^{2}$
with $\gamma=1.40$ for air, and $C(0)=\gamma V/\rho c^{2}$.
Integrating $h(\omega)/\omega^{2}$ around the real axis indented above
the origin and closed by a large upper semicircle (on which
$h\sim\omega^{-1}$, so the arc contributes nothing) and taking real parts
gives the strict, dc-anchored invariant
\begin{equation}
  \Buddc\;\equiv\;\int_{0}^{\infty}\frac{\TL(f)}{f^{2}}\,\mathrm{d}f
  \;=\;\frac{20}{\ln 10}\,\frac{\pi^{2}\gamma V}{2\,c\,S}
  \quad[\mathrm{dB\,s}].
  \label{eq:budget}
\end{equation}
The isothermal excess $(\gamma-1)$ is, however, deposited entirely below
the audible band: the thermal transition of an unfilled cavity of
hydraulic radius \SI{1}{\milli\metre} lies near \SI{7}{\hertz}
(Sec.~\ref{sec:thermal}). The resource actually available to a designer
across the audio band is therefore the adiabatic budget
\begin{equation}
  \Budad\;=\;\frac{20}{\ln 10}\,\frac{\pi^{2}V}{2\,c\,S}
  \;\simeq\;42.9\,\frac{V}{cS}\quad[\mathrm{dB\,s}],
  \qquad \Budad=\Buddc/\gamma,
  \label{eq:budget_ad}
\end{equation}
and it is $\Budad$ that governs every audio-band statement below unless
$\Buddc$ is named explicitly. Section~\ref{sec:thermal} shows that part of
the stranded $(\gamma-1)$ excess can in fact be moved into the audio band
by filling the cavities, recovering usable budget without added envelope.
At room conditions ($c=\SI{343}{\metre\per\second}$) the adiabatic budget
reduces to a rule that can be evaluated by inspection,
\begin{equation}
  \Budad\;[\mathrm{dB\,ms}]\;=\;0.125\times(V/S)\;[\mathrm{mm}],
  \label{eq:rule}
\end{equation}
so the entire audio-band resource is set by one length: the cavity volume
divided by the bore area. Figure~\ref{fig:fig1}(a) maps
Eq.~(\ref{eq:rule}) over the design plane and locates the three
representative classes of Table~\ref{tab:forms}.

\begin{table}[t]
  \centering\small
  \caption{Representative device classes. No shell geometry is assumed;
  a class is fixed by its total cavity volume $V$, the number of cavities
  $N$ it is divided into, and the open bore area $S$. The tabulated
  budget is the audio-band value $\Budad$ of Eq.~(\ref{eq:rule}).}
  \label{tab:forms}
  \begin{tabular}{llrrrrr}
    \toprule
    & class & $V$ (\si{\cubic\milli\metre}) & $N$ &
    bore $\varnothing$ (\si{\milli\metre}) & $S$ (\si{\square\milli\metre}) &
    $\Budad$ (\si{\decibel\,\milli\second})\\
    \midrule
    F1 & canal plug            &  150 & 3 & 2.0 & 3.14 & \phantom{0}5.97\\
    F2 & deep canal plug       &  400 & 4 & 2.0 & 3.14 & 15.91\\
    F3 & concha+canal mould    & 1200 & 6 & 2.0 & 3.14 & 47.73\\
    \bottomrule
  \end{tabular}
\end{table}

Equation~(\ref{eq:budget_ad}) is the working statement of this paper. The
resonators can place transmission loss anywhere in the spectrum, but the
$f^{-2}$-weighted area under $\TL(f)$ is fixed by $V$ and $S$ alone.
Figure~\ref{fig:fig1}(b,c) verifies this numerically: four structurally
dissimilar allocations of the same \SI{400}{\cubic\milli\metre}, a
three-resonator array tuned to \SIlist{220;300;380}{\hertz}, a single
resonator at \SI{1.5}{\kilo\hertz}, a six-resonator ladder spanning
\SIrange{0.6}{3.9}{\kilo\hertz}, and a four-resonator short-neck array
across \SIrange{2}{6}{\kilo\hertz}, were evaluated by transfer-matrix
analysis with Crandall viscothermal neck impedances~\cite{crandall1926}.
Their transmission-loss spectra differ completely, yet their running
integrals converge to the same $\Budad$ within
\SIrange{0.0}{0.8}{\percent}.

\subsection{Which protectors the budget binds}
\label{sec:scope}

The derivation requires $t(0)=1$: the device must pass zero frequency.
This is not a technicality: stating it removes an apparent contradiction
with experiment. An occluding earplug has $t(0)=0$, the integral
$\int\ln|t|^{-1}\omega^{-2}\mathrm{d}\omega$ diverges, and
Eq.~(\ref{eq:budget}) places no constraint on it. The recently reported
low-frequency gains of non-vented meta-earplugs~\cite{carillo2026} are
obtained in that regime: the resonators there reshape the reflection
coefficient at the plug's medial surface, altering the standing-wave field
in the residual ear-canal
cavity~\cite{zurbrugg2014occlusion,porschmann2000ownvoice}, rather than
filtering transmission along an open channel. The two mechanisms are
topologically distinct, a one-port (reflective) versus a two-port
(transmitting) problem; this distinction itself sets different bandwidth
limits~\cite{lee2019damped}, and the numbers reported here neither
contradict nor bound those results.

The same observation cuts the other way, and is the more useful statement
of the sum rule for device work: in sum-rule terms, occlusion is free, and
what the budget prices is selective transparency.
A protector that simply blocks (a foam plug, a solid mould, a resistive
element in series with the bore) is not budget-limited, because it is
not transparent at dc. This is why uniform-attenuation musicians'
earplugs~\cite{killion1988}, which combine an acoustic resistance with a
small cavity, achieve a broadly flat \SIrange{15}{20}{\decibel} in a
volume far smaller than Eq.~(\ref{eq:budget_ad}) would seem to allow: their
attenuation is dissipative and broadband, not reactive and selective.
Any frequency-selective concept must be shown to beat that baseline, and
we return to the comparison in Sec.~\ref{sec:atlas}.

\subsection{The bore, not the volume, is the cheap lever}

Since $\Budad\propto S^{-1}\propto d^{-2}$, halving the bore diameter
quadruples the budget. At $V=\SI{400}{\cubic\milli\metre}$ the budget
runs \SIlist{28.3;15.9;10.2;7.1}{\decibel\,\milli\second} for bore
diameters of \SIlist{1.5;2.0;2.5;3.0}{\milli\metre}: a factor of four
from a dimension that costs no packaging space at all. The limit is that
a sufficiently narrow bore ceases to be a transparent channel and becomes
a broadband acoustic resistance, at which point the device has drifted
into the uniform-attenuation class of Sec.~\ref{sec:scope} and left the
scope of Eq.~(\ref{eq:budget_ad}). For the \SIrange{1}{4}{\milli\metre}
range considered here the bore's own viscous loss stays below
\SI{1}{\decibel} across the audio band and the device remains
dc-transparent.

\section{Three constraints on spending the budget}
\label{sec:walls}

The budget is a ceiling, not a construction. Three further constraints
decide whether it can be spent where a designer wants it. The first two
are fundamental, one causal and one viscous; the third, the flanking
path, is a representative engineering ceiling rather than a fundamental
bound, and we label it as such in Sec.~\ref{sec:flank}.

\subsection{Constraint 1: the $f^{-2}$ kernel}

For a flat $\TL_{0}$ over $[f_{1},f_{2}]$,
\begin{equation}
  \TL_{0}^{\max}=\Budad\left(\frac{1}{f_{1}}-\frac{1}{f_{2}}\right)^{-1},
  \label{eq:flat}
\end{equation}
so at fixed fractional bandwidth the ceiling scales as the centre
frequency: one octave at \SI{250}{\hertz} costs four times as much budget
as one octave at \SI{1}{\kilo\hertz} (Table~\ref{tab:ceilings}). Inverting
the relation makes the cost concrete: reaching a flat \SI{10}{\decibel}
through a \SI{2.0}{\milli\metre} bore requires \SI{1047}{\cubic\milli\metre}
of cavity over \SIrange{150}{400}{\hertz}, \SI{220}{\cubic\milli\metre}
over \SIrange{1}{8}{\kilo\hertz}, and only \SI{63}{\cubic\milli\metre} over
\SIrange{2}{4}{\kilo\hertz}, so the same ten decibels costs seventeen
times more volume at low frequency than in the band where noise-induced
damage is most readily accrued. The same volume-versus-low-frequency
tension constrains resonant absorbers generally, where the resonant
frequency falls with back volume~\cite{berchtenbreiter2024mpp}.

\begin{table}[t]
  \centering\small
  \caption{Flat transmission loss sustainable on the audio-band budget
  $\Budad$, Eq.~(\ref{eq:flat}). Entries in italics exceed the flanking
  floor of Sec.~\ref{sec:flank} and are not physically realisable; they
  indicate only that the causal budget has ceased to be the binding
  constraint.}
  \label{tab:ceilings}
  \begin{tabular}{lrrr}
    \toprule
    band & F1 & F2 & F3\\
    \midrule
    \SIrange{150}{400}{\hertz}    & \phantom{0}1.4 & \phantom{0}3.8 & \phantom{0}11.5\\
    \SIrange{300}{1000}{\hertz}   & \phantom{0}2.6 & \phantom{0}6.8 & \phantom{0}20.5\\
    \SIrange{1}{8}{\kilo\hertz}   & \phantom{0}6.8 & 18.2 & \textit{54.6}\\
    \SIrange{2}{6}{\kilo\hertz}   & 17.9 & \textit{47.7} & \textit{143.2}\\
    \SIrange{2}{4}{\kilo\hertz}   & 23.9 & \textit{63.6} & \textit{190.9}\\
    \SIrange{3}{6}{\kilo\hertz}   & 35.8 & \textit{95.5} & \textit{286.4}\\
    \bottomrule
  \end{tabular}
\end{table}

\subsection{Constraint 2: viscous accessibility}

Placing a resonance at $f_{0}$ inside a fixed cavity $V_{i}$ requires a
neck of area $A_{i}=(2\pi f_{0}/c)^{2}V_{i}L_{\mathrm{eff}}$, so low
$f_{0}$ forces a narrow neck, while the viscous boundary-layer thickness
$\delta_{v}=\sqrt{2\mu/\rho\omega}$ grows as $f_{0}$ falls. The two meet
and the resonance is extinguished. Table~\ref{tab:visc} makes this
concrete. Below about \SI{800}{\hertz} the required neck radius drops
below $\delta_{v}$ and the attainable peak transmission loss collapses by
two orders of magnitude, in every form factor. This is the mechanism
behind the \SIrange{220}{380}{\hertz} design of Fig.~\ref{fig:fig1}: its
budget is not lost, but it is deposited below \SI{100}{\hertz}, where it is
acoustically useless. Figure~\ref{fig:fig2}(a) plots constraints~1 and~2
together. They are independent in origin, one from causality and one from
viscosity, but they close on the same region of the spectrum.

\begin{table}[t]
  \centering\small
  \caption{Viscous accessibility. $a_{\mathrm{req}}$ is the neck radius
  that places the resonance of one F2 cavity ($V/N=\SI{100}{\cubic\milli\metre}$)
  at $f_{0}$, computed with the flanged effective length
  $L_{\mathrm{eff}}=L+1.7a_{\mathrm{req}}$ for a geometric neck length
  $L=\SI{2}{\milli\metre}$ (solved self-consistently); $\delta_{v}$ is the
  viscous boundary-layer thickness there. The last three columns give the
  peak transmission loss attainable from a single cavity of $V/N$ in each
  class.}
  \label{tab:visc}
  \begin{tabular}{rrrrrrr}
    \toprule
    $f_{0}$ (Hz) & $a_{\mathrm{req}}$ (\si{\micro\metre}) &
    $\delta_{v}$ (\si{\micro\metre}) & $a_{\mathrm{req}}/\delta_{v}$ &
    F1 (dB) & F2 (dB) & F3 (dB)\\
    \midrule
     150 &  22 & 179 & 0.12 & 0.00 & 0.00 & \phantom{0}0.01\\
     300 &  45 & 127 & 0.35 & 0.01 & 0.02 & \phantom{0}0.10\\
     500 &  75 &  98 & 0.77 & 0.05 & 0.20 & \phantom{0}0.79\\
     800 & 123 &  78 & 1.58 & 0.32 & \phantom{0}1.28 & \phantom{0}4.38\\
    1200 & 189 &  63 & 2.98 & \phantom{0}1.57 & \phantom{0}4.96 & 11.52\\
    2000 & 331 &  49 & 6.74 & \phantom{0}6.84 & 14.18 & 23.67\\
    3000 & 528 &  40 & 13.2 & 14.43 & 23.77 & 33.72\\
    5000 & 992 &  31 & 31.9 & 25.82 & 33.98 & 39.06\\
    \bottomrule
  \end{tabular}
\end{table}

\subsection{Constraint 3: the flanking path}
\label{sec:flank}

Equation~(\ref{eq:t}) assumes the bore is the only route to the eardrum.
No real device satisfies this. Transmission through the plug body, leakage
around the seal and bone conduction form a parallel path whose insertion
loss saturates in the range \SIrange{35}{50}{\decibel} for well-fitted
protectors~\cite{berger2003,berger1998development}, a ceiling also seen in
field measurements of impulse peak insertion
loss~\cite{murphy2012firearm} and in double-protection
studies~\cite{luan2022double}. Treating the two paths as incoherent, the
measured attenuation is the power combination
\begin{equation}
  \TL_{\mathrm{tot}}=-10\log_{10}\bigl(10^{-\TL_{\mathrm{bore}}/10}
    +10^{-\TL_{\mathrm{fl}}/10}\bigr),
  \label{eq:flank}
\end{equation}
so once $\TL_{\mathrm{bore}}$ approaches $\TL_{\mathrm{fl}}$, additional
budget buys almost nothing: with $\TL_{\mathrm{fl}}=\SI{40}{\decibel}$,
raising the bore's contribution from \SIrange{25}{40}{\decibel} moves the
total from \SIrange{24.9}{37.0}{\decibel}, and from
\SIrange{40}{55}{\decibel} moves it only from
\SIrange{37.0}{39.9}{\decibel}.
We adopt $\TL_{\mathrm{fl}}=\SI{40}{\decibel}$ throughout as a
representative value. Equation~(\ref{eq:flank}) is a phenomenological,
incoherent-path model, not a fundamental causal bound: if the bore and
flanking paths are partially coherent their relative phase enters, and the
true flanking limit is frequency dependent rather than the flat
\SI{40}{\decibel} used here. The qualitative conclusion, that the
design space is closed from above by a mechanism the scattering bounds do
not see, is insensitive to these details.
This flanking ceiling is why the italicised entries of
Table~\ref{tab:ceilings} should be read as ``the budget is no longer
binding here,'' not as achievable attenuation, and it is the reason that
enlarging the form factor produces diminishing returns long before the
budget is exhausted.

\subsection{A recoverable margin: thermal stranding}
\label{sec:thermal}

\begin{figure}[t]
  \centering
  \includegraphics[width=\linewidth]{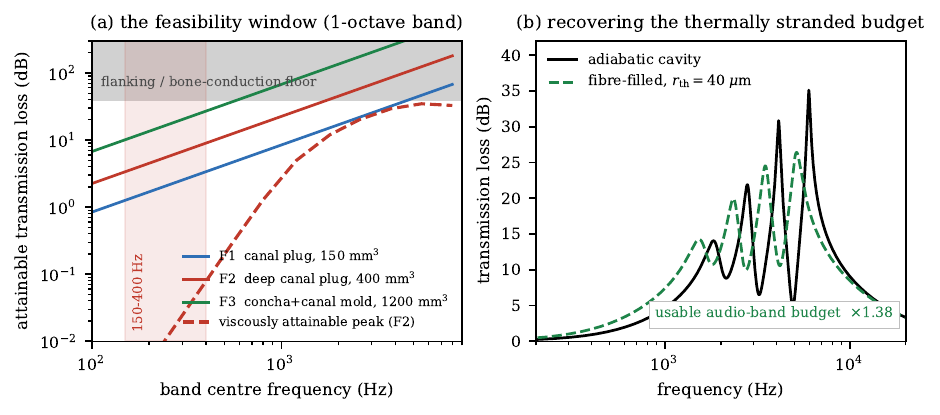}
  \caption{(a) Constraints 1--3. Solid: the causal ceiling of
  Eq.~(\ref{eq:flat}) for a one-octave band, for the three classes of
  Table~\ref{tab:forms}. Dashed: the viscously attainable peak
  transmission loss of a single F2 cavity. Shaded: the flanking floor of
  Sec.~\ref{sec:flank}. The \SIrange{150}{400}{\hertz} band targeted by
  low-frequency-selective designs lies beneath both lower constraints.
  (b) Recovering the thermally stranded budget: transmission loss of the
  same four-resonator array with adiabatic cavities and with fibre-filled
  cavities of effective thermal radius \SI{40}{\micro\metre}. The strict
  dc-anchored budget is $\Buddc=\gamma\Budad$; filling converts part of
  that excess from inaudible frequencies into the audio band.}
  \label{fig:fig2}
\end{figure}

As noted at Eq.~(\ref{eq:budget}), the sum rule is anchored on the
static compliance of the side branches, and static compression in a
thermally conducting cavity is isothermal, giving
$C(0)=\gamma V/\rho c^{2}$ and the strict dc invariant
$\Buddc=\gamma\Budad\approx1.40\,\Budad$.
In an unfilled cavity of hydraulic radius \SI{1}{\milli\metre} the thermal
transition lies near \SI{7}{\hertz}, so the entire $(\gamma-1)$ excess is
deposited below the audible range, stranded in the same way, and for
the same kind of reason, that viscosity strands the low-frequency budget.
Filling the cavities with a fibrous or lattice medium of effective thermal
radius \SI{40}{\micro\metre} moves the transition to about
\SI{4}{\kilo\hertz}, and returns a factor of \num{1.38} in usable
audio-band budget, recovering most of the $(\gamma-1)$ excess and raising
the effective $\Budad$ towards $\Buddc$ [Fig.~\ref{fig:fig2}(b)].
The mechanism is the same one that makes damping material raise the
effective volume of a loudspeaker enclosure~\cite{zwikker1949}; to our
knowledge it has not been exploited in resonator-based hearing protection,
and it costs no envelope.

\section{Allocating the budget}
\label{sec:alloc}

\subsection{Hazard-weighted water-filling}

Given a source with spectral density $P(f)$ and a hazard weighting
$W(f)$, the weighted transmitted energy is
$E_{t}=\int W(f)P(f)e^{-2u(f)}\mathrm{d}f$ with $u=\ln|t|^{-1}$.
Minimising $E_{t}$ subject to $\int uf^{-2}\mathrm{d}f=\Budad\ln 10/20$
and $u\ge0$ gives
\begin{equation}
  u^{\star}(f)=\frac{1}{2}\bigl[\ln\bigl(f^{2}W(f)P(f)/\theta\bigr)\bigr]_{+},
  \label{eq:waterfill}
\end{equation}
a water-filling solution in which attenuation is poured into the spectrum
wherever the value density $f^{2}W(f)P(f)$ rises above a level set
by the budget; causality-guided allocation of an absorption spectrum in
this spirit has been developed for micro-perforated
partitions~\cite{bravo2022causally}. The $f^{2}$ factor is the price
implied by the $f^{-2}$
kernel of Eq.~(\ref{eq:budget_ad}) and it is not a small correction: it
displaces the optimum well above the source's spectral peak.

We take $W$ to be A-weighting, which is the coarsest defensible proxy for
the frequency dependence of auditory hazard. We stress what this does
not claim. Noise-induced hearing loss is a function of level,
duration and spectrum jointly, and is not confined to any single band;
the familiar \SI{4}{\kilo\hertz} audiometric notch arises largely from
ear-canal and middle-ear transfer rather than from source
spectra~\cite{natarajan2023noise}. The
role of $W$ here is only to make explicit that the value of a decibel of
attenuation is frequency dependent, and to place that dependence where a
designer can substitute a better weighting; a validated damage-risk
weighting such as the auditory hazard assessment algorithm for
humans~\cite{price1996evaluation,price2007validation}, now embodied in
military noise-limit standards~\cite{milstd1474e}, would enter
Eq.~(\ref{eq:waterfill}) in exactly the same place.

The conclusions below do not depend on this choice.
Repeating the analysis with a damage-risk-like weighting (an outer- and
middle-ear pressure transfer peaking near \SI{3.5}{\kilo\hertz}, the
physical origin of the \SI{4}{\kilo\hertz} audiometric notch) in place of
A-weighting moves the value-density peak for the
\SI{0.3}{\milli\second} blast from \SI{2.21}{\kilo\hertz} to
\SI{2.68}{\kilo\hertz} and lowers the transparency ceiling $R_{\infty}$
(Sec.~\ref{sec:exchange}) from \SI{0.51}{\decibel} to \SI{0.33}{\decibel}
(Table~\ref{tab:sens}). A weighting that emphasises the damage band more
strongly therefore pushes the optimum further above the source peak
and tightens the transparency ceiling, not the reverse: these
qualitative results are properties of the source--speech overlap, robust
to the hazard weighting across the range a designer would plausibly adopt.

\begin{table}[t]
  \centering\small
  \caption{Robustness of the headline results to the hazard weighting
  $W(f)$, for the \SI{0.3}{\milli\second} blast. The damage-risk surrogate
  is an outer/middle-ear pressure transfer peaking near
  \SI{3.5}{\kilo\hertz} times the A-weighting skirt; its exact form is in
  the released code, so a reviewer may substitute a validated model. In
  every row the weighting moves the result in the same direction, so the
  conclusions do not depend on the choice.}
  \label{tab:sens}
  \begin{tabular}{lrrr}
    \toprule
    & unweighted & A-weighting & damage-risk\\
    \midrule
    value-density peak (\si{\kilo\hertz}) & 2.03 & 2.21 & 2.68\\
    $R_{\infty}$, $[300,3400]$\,\si{\hertz} (dB) & 0.71 & 0.51 & 0.33\\
    \bottomrule
  \end{tabular}
\end{table}

Three sources are carried through: a Friedlander blast of positive-phase
duration $T$ and shock rise time $\tau_{r}$, for which $P(f)\propto
f^{2}(f_{c}^{2}+f^{2})^{-2}[1+(f/f_{r})^{2}]^{-1}$ with $f_{c}=1/2\pi T$
and $f_{r}=1/2\pi\tau_{r}$, evaluated at $T=\SI{0.3}{\milli\second}$
(small-arms muzzle blast) and $T=\SI{1}{\milli\second}$ (larger calibre
or enclosed firing); and a continuous broadband machinery spectrum with a
$-\SI{6}{\decibel}$/octave roll-off above \SI{500}{\hertz} (a pink-noise
surrogate; the exact form is given in the released code).
Figure~\ref{fig:fig3}(a) shows the displacement the price signal
produces. For the \SI{0.3}{\milli\second} blast the source peaks at
\SI{525}{\hertz} but the hazard-weighted value density peaks at
\SI{2.21}{\kilo\hertz}; for the \SI{1}{\milli\second} blast, at
\SI{159}{\hertz} and \SI{1.86}{\kilo\hertz}. The most
cost-effective place to attenuate a muzzle blast lies roughly two octaves
above where its energy is concentrated.

\subsection{From a prohibition to an exchange rate}
\label{sec:exchange}

The premise of a selective protector is that speech remains usable. It is
tempting to impose this as $u=0$ on a speech window $\mathcal{W}$, and
doing so yields a bound that is independent of the budget entirely: since
energy inside $\mathcal{W}$ passes unattenuated by construction,
\begin{equation}
  R\;\le\;R_{\infty}\;=\;10\log_{10}
  \frac{\int_{0}^{\infty}WP\,\mathrm{d}f}{\int_{\mathcal{W}}WP\,\mathrm{d}f},
  \label{eq:overlap}
\end{equation}
independently of $V$, $S$, the resonator count and indeed of the
mechanism. For $\mathcal{W}=[\SI{300}{\hertz},\SI{3400}{\hertz}]$ and the
\SI{0.3}{\milli\second} blast, $R_{\infty}=\SI{0.70}{\decibel}$
unweighted and \SI{0.51}{\decibel} A-weighted; for the
\SI{1}{\milli\second} blast, \SI{2.78}{\decibel} and
\SI{0.63}{\decibel}; for continuous noise, \SI{3.38}{\decibel} and
\SI{1.74}{\decibel}. Hazard weighting does not rescue the strictly
transparent concept; it tightens it.

\begin{figure}[t]
  \centering
  \includegraphics[width=.62\linewidth]{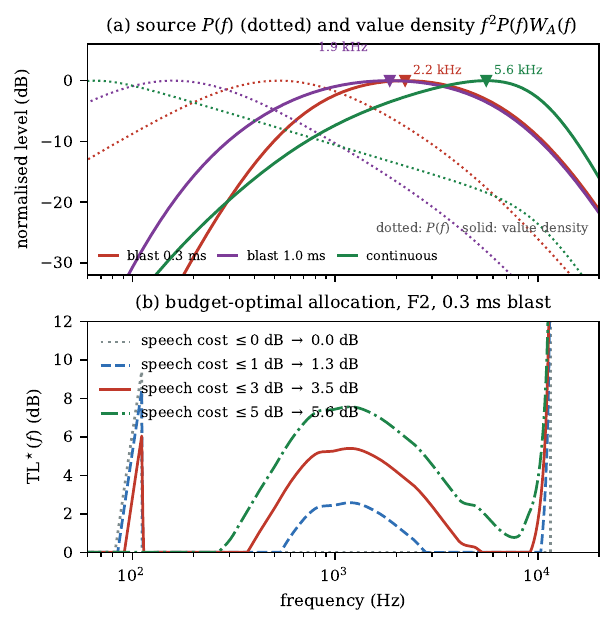}
  \caption{(a) Source spectra (dotted) and hazard-weighted value
  densities $f^{2}P(f)W_{A}(f)$ (solid), each normalised to its own
  maximum; triangles mark the value-density peaks. (b) Budget-optimal
  allocations from Eq.~(\ref{eq:twoconstraint}) for class F2 and the
  \SI{0.3}{\milli\second} blast, at four speech allowances; the legend
  gives the realised A-weighted protection. The narrow spikes at the
  edges of the speech-importance support are the residue of the strict
  constraint: budget accumulates where speech importance vanishes, and
  buys almost nothing there.}
  \label{fig:fig3}
\end{figure}

Equation~(\ref{eq:overlap}) is correct, and taken alone it reads as a
prohibition. It is also an artefact of a constraint that no real design
imposes. A protector that attenuates the speech range by a few decibels
uniformly loses little intelligibility, and in a high-level environment
may even improve it by raising the speech-to-threat ratio at the ear;
audibility above \SI{3}{\kilo\hertz}, where much of the damage energy also
sits, is what dominates intelligibility for the impaired
listener~\cite{moore2016review}. We
therefore replace the hard constraint by a continuous price. Let $i(f)$ be
the ANSI S3.5 speech-importance function~\cite{ansi1997}, read as a
density normalised so that $\int i\,\mathrm{d}f=1$, and define the
articulation-weighted speech cost
\begin{equation}
  \Cspeech=\int i(f)\,\TL(f)\,\mathrm{d}f\quad[\mathrm{dB}],
  \label{eq:cost}
\end{equation}
which equals $T$ for a uniform $\TL=T$ across the speech range and zero
for perfect transparency. We emphasise that $\Cspeech$ is an
articulation-weighted attenuation surrogate, not a direct measure
of intelligibility loss: it weights attenuation by band importance but is
linear in decibels, and in particular it under-prices narrow notches,
whose true cost in a speech-intelligibility-index or recognition sense can
be larger. We use it as a tractable, monotone proxy and return to its
limitations in Sec.~\ref{sec:disc}.
Maximising hazard-weighted protection subject
to both the budget and $\Cspeech\le\Cspeech^{\max}$ gives
\begin{equation}
 u^{\star}(f)=\frac12\Bigl[\ln\frac{2f^{2}W(f)P(f)}
    {\lambda+\mu\,\kappa\,i(f)f^{2}}\Bigr]_{+},
  \qquad\kappa=20/\ln 10,
  \label{eq:twoconstraint}
\end{equation}
with $\lambda$ set by the budget and $\mu$ by the speech allowance
(a derivation from the two-constraint stationarity conditions is given in
the released code).
Figure~\ref{fig:fig3}(b) shows the resulting allocations. At
$\Cspeech^{\max}=0$ the optimum can only pile attenuation outside the
support of $i(f)$ (below \SI{140}{\hertz} and above
\SI{9}{\kilo\hertz}) where the blast carries almost no hazard-weighted
energy, and the realised protection is \SI{0.00}{\decibel}. Allowing even
\SI{1}{\decibel} of speech cost releases \SI{1.33}{\decibel}.

\subsection{The exchange rate, and where the budget stops it}

\begin{figure}[t]
  \centering
  \includegraphics[width=\linewidth]{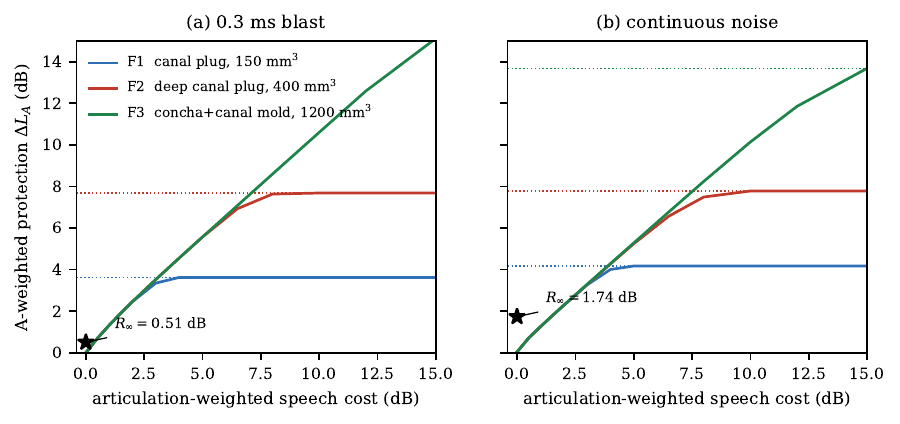}
  \caption{Protection--speech-cost Pareto fronts under
  Eq.~(\ref{eq:twoconstraint}) with the flanking floor of
  Eq.~(\ref{eq:flank}), for (a) the \SI{0.3}{\milli\second} blast and
  (b) continuous broadband noise. Stars mark the strict-transparency
  ceiling $R_{\infty}$ of Eq.~(\ref{eq:overlap}). Dotted lines mark the
  saturation set by each class's budget. The near-common initial slope is
  specific to the present source and speech-cost model (see text).}
  \label{fig:fig4}
\end{figure}

Figure~\ref{fig:fig4} is the central result. Sweeping
$\Cspeech^{\max}$ traces a Pareto front between protection and speech cost
for each class and source. Two features are worth isolating.

First, under the source and speech-cost model used here, the initial
slope is approximately the same for all three classes: about
\SI{1.3}{\decibel} of A-weighted protection per decibel of
articulation-weighted speech cost, for both impulsive and continuous
sources. This near-independence of form factor is a numerical property of
the present model rather than a constant derived from the sum rule: the
slope is governed by the overlap between the threat spectrum $W(f)P(f)$
and the speech-importance function $i(f)$, and would shift under a
different source, hazard weighting or importance function. Within these
inputs, near the transparent limit the exchange rate is a property of that
overlap, not of the device.

Second, the classes differ only in where the front saturates: at
\SI{3.6}{\decibel} for F1, \SI{7.7}{\decibel} for F2 and
\SI{12.7}{\decibel} for F3 against the \SI{0.3}{\milli\second} blast.
That saturation is the volume budget, modulated by the flanking floor.
The two design variables are therefore cleanly separated:
the speech allowance determines how far along the front a
design sits, and the volume budget determines where the front ends.
The resonator count and layout determine neither; they determine only how
closely a realisable device approaches the front, which is the subject of
the next section.

\section{A design atlas}
\label{sec:atlas}

\begin{figure}[t]
  \centering
  \includegraphics[width=\linewidth]{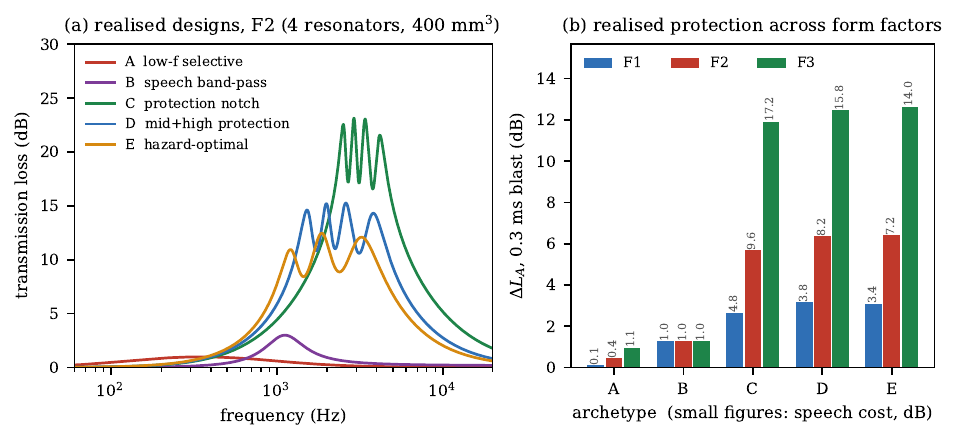}
  \caption{(a) Realised transmission loss of the five archetypes in class
  F2 (four resonators, \SI{400}{\cubic\milli\metre}), including the
  flanking path; the dotted line is the flanking floor. (b) Realised
  A-weighted protection against the \SI{0.3}{\milli\second} blast for
  each archetype and class; the small figures above each bar give the
  articulation-weighted speech cost incurred.}
  \label{fig:fig5}
\end{figure}

We now place five commercially motivated archetypes inside this space and
synthesise, for each, a realisable array of $N$ Helmholtz resonators
sharing the class volume. Neck radii, the volume split and an added mesh
resistance per branch are optimised; the resulting transmission loss is
computed by transfer-matrix analysis with Crandall neck impedances and
combined with the flanking path through Eq.~(\ref{eq:flank}). The
archetypes are: \textbf{A}, low-frequency selective, with resonances
constrained to \SIrange{150}{400}{\hertz}; \textbf{B}, speech band-pass,
constrained to $\Cspeech\le\SI{1}{\decibel}$; \textbf{C}, a protective
notch confined to \SIrange{2}{6}{\kilo\hertz}; \textbf{D}, mid- and
high-frequency protection across \SIrange{1}{8}{\kilo\hertz}; and
\textbf{E}, the unconstrained hazard-matched optimum.
Table~\ref{tab:atlas} and Fig.~\ref{fig:fig5} give the results.

\begin{table}[t]
  \centering\small
  \caption{Design atlas. $\Delta L_{A}$ is the A-weighted energy
  reduction against the \SI{0.3}{\milli\second} blast and against
  continuous noise; $\Cspeech$ is the articulation-weighted speech cost
  of Eq.~(\ref{eq:cost}). All entries are realised finite-$Q$ designs
  including the flanking path.}
  \label{tab:atlas}
  \begin{tabular}{llrrrr}
    \toprule
    class & archetype & peak $\TL$ (dB) & $\Delta L_{A}$ blast &
    $\Delta L_{A}$ cont. & $\Cspeech$ (dB)\\
    \midrule
    F1 & A low-frequency selective & \phantom{0}0.1 & 0.11 & 0.09 & \phantom{0}0.10\\
    F1 & B speech band-pass        & \phantom{0}2.7 & 1.30 & 0.88 & \phantom{0}1.02\\
    F1 & C protection notch        & 14.9 & 2.66 & 3.44 & \phantom{0}4.78\\
    F1 & D mid+high protection     & \phantom{0}8.9 & 3.16 & 3.17 & \phantom{0}3.82\\
    F1 & E hazard-optimal          & \phantom{0}7.7 & 3.10 & 2.89 & \phantom{0}3.39\\
    \midrule
    F2 & A low-frequency selective & \phantom{0}1.0 & 0.46 & 0.34 & \phantom{0}0.45\\
    F2 & B speech band-pass        & \phantom{0}3.0 & 1.28 & 0.89 & \phantom{0}1.02\\
    F2 & C protection notch        & 23.1 & 5.70 & 6.53 & \phantom{0}9.60\\
    F2 & D mid+high protection     & 15.2 & 6.36 & 6.36 & \phantom{0}8.20\\
    F2 & E hazard-optimal          & 12.5 & 6.42 & 5.95 & \phantom{0}7.20\\
    \midrule
    F3 & A low-frequency selective & \phantom{0}3.4 & \phantom{0}0.96 & \phantom{0}0.71 & \phantom{0}1.08\\
    F3 & B speech band-pass        & \phantom{0}3.1 & \phantom{0}1.26 & \phantom{0}0.88 & \phantom{0}1.02\\
    F3 & C protection notch        & 31.6 & 11.91 & 12.03 & 17.22\\
    F3 & D mid+high protection     & 24.6 & 12.46 & 11.93 & 15.80\\
    F3 & E hazard-optimal          & 20.0 & 12.64 & 11.43 & 13.98\\
    \bottomrule
  \end{tabular}
\end{table}

Four readings follow.

Archetype A gains almost nothing in the vented class, and enlarging the
device does not rescue it. Constrained to its nominal
\SIrange{150}{400}{\hertz} target,
the best three-resonator canal plug reaches \SI{0.1}{\decibel} of peak
transmission loss and \SI{0.11}{\decibel} of A-weighted protection. An
eightfold volume increase to a full concha mould raises this only to
\SI{3.4}{\decibel} peak and \SI{0.96}{\decibel} of protection. Both lower
constraints bind at once: the budget is expensive here and viscosity makes
even the budgeted attenuation unreachable. A device aiming at this band is
not underperforming; it is operating in the one region where every
constraint in Sec.~\ref{sec:walls} binds simultaneously.

Archetype B saturates and is insensitive to volume. A strictly
speech-transparent band-pass returns \SIrange{1.26}{1.30}{\decibel}
across a factor of eight in cavity volume, the behaviour
Eq.~(\ref{eq:overlap}) predicts. The threat and the speech range overlap
too strongly for spectral separation alone to deliver protection against
small-arms blast. It fares better against longer-duration sources, where
$R_{\infty}$ rises.

Archetypes C--E scale with the budget and are the viable
territory. Protection rises roughly in proportion to $\Budad$, from
\SIrange{3.1}{3.2}{\decibel} in a \SI{150}{\cubic\milli\metre} plug to
\SIrange{12.5}{12.6}{\decibel} in a \SI{1200}{\cubic\milli\metre} mould,
and viscosity is not a constraint in this band (Table~\ref{tab:visc}).
Crucially, the same designs work against continuous noise as well as
against blast, which widens the commercial case considerably.

Do not notch narrowly. Archetype E delivers slightly more
protection than the narrow \SIrange{2}{6}{\kilo\hertz} notch C at
appreciably lower speech cost in every class: for F2, \num{6.42} against
\SI{5.70}{\decibel} of protection at a speech cost of \num{7.20} against
\SI{9.60}{\decibel}. Attenuation should be spread along the value density
$f^{2}W(f)P(f)$ rather than concentrated on a target band chosen by
analogy with the audiometric notch. This is the practical content of the
$f^{2}$ price signal. The margin also reflects that the linear speech cost
$\Cspeech$ under-prices the narrow notch; a full intelligibility model
would widen it further in E's favour.

\section{Discussion}
\label{sec:disc}

Read as a resource-allocation problem, the design space has a simple
structure. The resource is one length, $V/S$, and the bore is the
cheaper of the two ways to change it. Three constraints bound how the
resource can be spent: causality prices low-frequency selectivity through
the $f^{-2}$ kernel, viscosity makes it unreachable at earplug scale, and
a flanking path caps the useful range from above at a level the causal
bounds do not see. Within those constraints, the allocation is a
water-filling problem against $f^{2}W(f)P(f)$, and the transparency
requirement is not a constraint but a price, whose exchange rate near the
transparent limit is, for the source and speech-cost model used here,
a property of the threat rather than of the device.

The prescription that follows is specific. Spend the budget where the
value density peaks, roughly \SIrange{1.5}{3}{\kilo\hertz} for small-arms
blast and higher for continuous machinery noise, and spread rather than
notch. Choose the bore before choosing the volume. Fill the cavities.
Budget the intelligibility loss explicitly (a few decibels of
articulation-weighted attenuation is the difference between a device that
does nothing and one that delivers useful protection) and verify it
psychoacoustically rather than assuming a transparent window. Regard the
resonator array as the last variable, not the first.

Two relationships to existing work deserve emphasis. Our
Eq.~(\ref{eq:budget}) is a specialisation of the scattering identities of
Refs.~\cite{norris2018,meng2022}; the contemporaneous general sum rule of
Ref.~\cite{qu2026} arrives at the same $\omega^{-2}$ kernel from
extinction cross-sections and static effective properties, and reaches a
parallel constructive conclusion, that the conserved integral can be
reshaped, though not enlarged, by appropriate spectral design. The
side branches here are Fano-type scatterers~\cite{fano1961effects}, and
the two-port extension of that sum rule~\cite{qu2025generalized} shows the
same causal budget governs their absorption as well as their transmission,
so the ceilings below are not artefacts of treating transmission alone.
The present work adds the device-variable statement, the two constraints
that scattering theory does not contain, and the allocation problem. And,
as Sec.~\ref{sec:scope} makes explicit, none of it constrains occluding
protectors, whose low-frequency performance is governed by the medial
reflection coefficient rather than by transmission along an open
channel~\cite{carillo2026}. The same vented, transfer-matrix description
underlies Helmholtz-resonator silencers for
ducts~\cite{gao2024silencer,bricault2022silencer},
where the branch reshapes transmission rather than the residual-cavity
standing wave that governs the occluded ear~\cite{brummund2021occlusion}.

Several limitations bound these conclusions. The lumped model of
Eq.~(\ref{eq:t}) assumes sub-wavelength branches and a single junction.
The $f^{-2}$ kernel of the sum rule is robust to relaxing this (it
follows from the low-frequency analyticity of $h(\omega)$ and survives in
distributed and multi-junction layouts), but the coefficient is not:
the clean $\Budad=0.125\,V/S$ relation is specific to the single-junction,
lumped-branch geometry, and distributed propagation changes the constant
relating the budget to $V/S$, as multi-port scattering-matrix analyses of
duct networks make explicit~\cite{calmettes2023multiport}.
The speech cost of Eq.~(\ref{eq:cost}) is linear in decibels and is an
articulation-weighted surrogate rather than a true intelligibility metric;
it under-prices very narrow notches, and a full speech-intelligibility-index
calculation with a specified masking condition would be the next
refinement, as would substituting a validated damage-risk weighting for
A-weighting.
We treat energy rather than peak pressure, which is the quantity
Parseval's theorem makes tractable but not the only relevant hazard
metric, and we exclude the occlusion-effect pathway, which sets its own
floor and which is precisely where resonator arrays have already proved
themselves~\cite{carillo2022,carillo2023,carillo2025}. Most importantly,
the entire analysis is linear and time-invariant. Muzzle blast at the ear
reaches levels at which orifice flow becomes
nonlinear~\cite{ingardising1967,cavalieri2024pressure}, and
amplitude-dependent resistance
breaks the assumptions behind Eqs.~(\ref{eq:budget_ad})
and~(\ref{eq:overlap}) alike. Since level dependence, unlike frequency
dependence, does not require speech and threat to be spectrally
separable, it is the one route to protection that the constraints
described here do not close, and the atlas above quantifies what is
left on the table by staying linear.

\section{Conclusion}

For a vented, dc-transparent hearing protector, the $f^{-2}$-weighted
integral of transmission loss is fixed by the ratio of total cavity
volume to bore area. The strict dc invariant is
$\Buddc=(20/\ln10)\,\pi^{2}\gamma V/2cS$; because the isothermal excess is
stranded below the audible band, the resource available across the audio
band is $\Budad=\Buddc/\gamma$, equal to $0.125\,V/S$ in
decibel-milliseconds when $V/S$ is expressed in millimetres.
Occluding protectors are not bound by this relation. Interpreting the
budget as a resource and allocating it against a hazard-weighted value
density gives a water-filling rule whose $f^{2}$ price signal favours
frequencies well above the source's spectral peak. Replacing strict
speech transparency by a continuous articulation-weighted price converts a
volume-independent prohibition into an exchange rate, roughly
\SI{1.3}{\decibel} of protection per decibel of speech cost under the
source and speech-cost model used here, with the volume budget setting
only where that exchange saturates.
Across three form factors, low-frequency-selective designs in the vented
class stay near a decibel, while hazard-matched mid- and high-frequency
designs reach twelve. This does not place low-frequency protection out of
reach: occluding and reflective designs obtain it by a route the budget
does not govern (Sec.~\ref{sec:scope}); within the vented, dc-transparent
class it is simply the costliest place to spend a fixed volume budget. The
value of the sum rule for design is that it prices these choices in
advance.

\subsection*{Funding}
This work was supported by JST SPRING, Grant Number JPMJSP2119.

\subsection*{Conflict of interest}
The authors declare no conflict of interest.

\subsection*{Author contributions}
G.H. led and supervised the study and developed the theory. J.L.
contributed to the conceptualisation and provided audiological and
ergonomic input. H.K. and H.N. implemented the transfer-matrix and
allocation computations and prepared the figures. All authors discussed
the results and wrote the manuscript.

\subsection*{Data availability statement}
The transfer-matrix model, sum-rule verification, allocation solver, the
two-constraint derivation, the explicit continuous-source spectrum, and the
archetype-synthesis code that reproduce all figures and tables in this
article are available in Zenodo at
\texttt{https://doi.org/10.5281/zenodo.22684722}.

\bibliographystyle{unsrtnat}
\bibliography{nplug_GH_reviewed}

@article{carillo2022,
  author      = {Carillo, K{\'e}vin and Sgard, Franck and Dazel, Olivier and Doutres, Olivier},
  title       = {Reduction of the occlusion effect induced by earplugs using quasi perfect broadband absorption},
  journal     = {Scientific Reports},
  volume      = {12},
  number      = {1},
  pages       = {15336},
  year        = {2022},
  month       = {Sep},
  issn        = {2045-2322},
  doi         = {10.1038/s41598-022-19641-3},
  url         = {https://doi.org/10.1038/s41598-022-19641-3}
}

@article{carillo2023,
  author      = {Carillo, K{\'{e}}vin and Sgard, Franck and Dazel, Olivier and Doutres, Olivier},
  title       = {Passive earplug including {Helmholtz} resonators arranged in series to achieve broadband near zero occlusion effect at low frequencies},
  journal     = {The Journal of the Acoustical Society of America},
  volume      = {154},
  number      = {4},
  pages       = {2099--2111},
  year        = {2023},
  month       = {Oct},
  issn        = {0001-4966},
  doi         = {10.1121/10.0021185},
  url         = {https://doi.org/10.1121/10.0021185}
}

@article{carillo2025,
  author      = {Carillo, K{\'{e}}vin and Sgard, Franck and Dazel, Olivier and Doutres, Olivier},
  title       = {Enhancing acoustic comfort for earplug users: objective and subjective evaluation of bone-conducted sound with meta-earplugs incorporating {Helmholtz} resonators},
  journal     = {Applied Acoustics},
  volume      = {240},
  pages       = {110929},
  year        = {2025},
  month       = {Dec},
  issn        = {0003-682X},
  doi         = {10.1016/j.apacoust.2025.110929},
  url         = {https://doi.org/10.1016/j.apacoust.2025.110929}
}

@article{carillo2026,
  author      = {Carillo, K{\'{e}}vin and Sgard, Franck and Dazel, Olivier and Doutres, Olivier},
  title       = {Improving low-frequency attenuation of passive earplugs using {Helmholtz} resonators},
  journal     = {The Journal of the Acoustical Society of America},
  volume      = {159},
  number      = {4},
  pages       = {3702--3712},
  year        = {2026},
  month       = {Apr},
  issn        = {1520-8524},
  doi         = {10.1121/10.0043161},
  url         = {https://doi.org/10.1121/10.0043161}
}

@article{jimenez2017,
  author      = {Jim{\'e}nez, No{\'e} and Romero-Garc{\'i}a, Vicent and Pagneux, Vincent and Groby, Jean-Philippe},
  title       = {Rainbow-trapping absorbers: Broadband, perfect and asymmetric sound absorption by subwavelength panels for transmission problems},
  journal     = {Scientific Reports},
  volume      = {7},
  number      = {1},
  pages       = {13595},
  year        = {2017},
  month       = {Oct},
  issn        = {2045-2322},
  doi         = {10.1038/s41598-017-13706-4},
  url         = {https://doi.org/10.1038/s41598-017-13706-4}
}

@article{romero2016,
  author      = {Romero-Garc{\'{i}}a, V. and Theocharis, G. and Richoux, O. and Pagneux, V.},
  title       = {Use of complex frequency plane to design broadband and sub-wavelength absorbers},
  journal     = {The Journal of the Acoustical Society of America},
  volume      = {139},
  number      = {6},
  pages       = {3395--3403},
  year        = {2016},
  month       = {Jun},
  issn        = {0001-4966},
  doi         = {10.1121/1.4950708},
  url         = {https://doi.org/10.1121/1.4950708}
}

@article{rozanov2000,
  author      = {Rozanov, K. N.},
  title       = {Ultimate thickness to bandwidth ratio of radar absorbers},
  journal     = {IEEE Transactions on Antennas and Propagation},
  volume      = {48},
  number      = {8},
  pages       = {1230--1234},
  year        = {2000},
  issn        = {0018-926X},
  doi         = {10.1109/8.884491},
  url         = {https://doi.org/10.1109/8.884491}
}

@article{yang2017,
  author      = {Yang, Min and Chen, Shuyu and Fu, Caixing and Sheng, Ping},
  title       = {Optimal sound-absorbing structures},
  journal     = {Materials Horizons},
  volume      = {4},
  number      = {4},
  pages       = {673--680},
  year        = {2017},
  issn        = {2051-6347},
  doi         = {10.1039/c7mh00129k},
  url         = {https://doi.org/10.1039/c7mh00129k}
}

@article{norris2018,
  author      = {Norris, Andrew N.},
  title       = {Integral identities for reflection, transmission, and scattering coefficients},
  journal     = {The Journal of the Acoustical Society of America},
  volume      = {144},
  number      = {4},
  pages       = {2109--2115},
  year        = {2018},
  month       = {Oct},
  issn        = {0001-4966},
  doi         = {10.1121/1.5058681},
  url         = {https://doi.org/10.1121/1.5058681}
}

@article{meng2022,
  author      = {Meng, Yang and Romero-Garc{\'{i}}a, Vicente and Gabard, Gw{\'{e}}na{\"{e}}l and Groby, Jean-Philippe and Bricault, Charlie and Goud{\'{e}}, S{\'{e}}bastien and Sheng, Ping},
  title       = {Fundamental constraints on broadband passive acoustic treatments in unidimensional scattering problems},
  journal     = {Proceedings of the Royal Society A: Mathematical, Physical and Engineering Sciences},
  volume      = {478},
  number      = {2265},
  pages       = {20220287},
  year        = {2022},
  month       = {Sep},
  issn        = {1364-5021},
  doi         = {10.1098/rspa.2022.0287},
  url         = {https://doi.org/10.1098/rspa.2022.0287}
}

@article{qu2026,
  author      = {Qu, Sichao and Yu, Zixiong and Dong, Erqian and Yang, Min and Fang, Nicholas X.},
  title       = {Acoustic Analogy of Quantum {Baldin} Sum Rule for Optimal Causal Scattering},
  journal     = {Phys. Rev. Lett.},
  volume      = {136},
  number      = {22},
  pages       = {226902},
  year        = {2026},
  month       = {Jun},
  publisher   = {American Physical Society},
  doi         = {10.1103/dbs8-g68w},
  url         = {https://link.aps.org/doi/10.1103/dbs8-g68w}
}

@article{bernland2011,
  author      = {Bernland, A and Luger, A and Gustafsson, M},
  title       = {Sum rules and constraints on passive systems},
  journal     = {Journal of Physics A: Mathematical and Theoretical},
  volume      = {44},
  number      = {14},
  pages       = {145205},
  year        = {2011},
  month       = {Apr},
  issn        = {1751-8113},
  doi         = {10.1088/1751-8113/44/14/145205},
  url         = {https://doi.org/10.1088/1751-8113/44/14/145205}
}

@book{crandall1926,
  author      = {Crandall, I. B.},
  title       = {Theory of Vibrating Systems and Sound},
  year        = {1926},
  publisher   = {Van Nostrand},
  address     = {New York}
}

@book{zwikker1949,
  author      = {Zwikker, C. and Kosten, C. W.},
  title       = {Sound Absorbing Materials},
  year        = {1949},
  publisher   = {Elsevier},
  address     = {Amsterdam}
}

@article{killion1988,
  author      = {Killion, M. C. and DeVilbiss, E. and Stewart, J.},
  title       = {An earplug with uniform 15-{dB} attenuation},
  journal     = {The Hearing Journal},
  volume      = {41},
  number      = {5},
  pages       = {14--17},
  year        = {1988}
}

@incollection{berger2003,
  author      = {Berger, Elliott H.},
  title       = {Hearing protection devices},
  booktitle   = {The Noise Manual},
  edition     = {Revised 5th},
  chapter     = {10},
  year        = {2003},
  publisher   = {American Industrial Hygiene Association},
  address     = {Fairfax, VA},
  editor      = {Berger, Elliott H. and Royster, Larry H. and Royster, Julia D. and Driscoll, Dennis P. and Layne, Martha},
  pages       = {379--454},
  isbn        = {9781931504027},
  doi         = {10.3320/978-1-931504-02-7.379},
  url         = {https://doi.org/10.3320/978-1-931504-02-7.379}
}

@techreport{ansi1997,
  author      = {{American National Standards Institute}},
  title       = {Methods for Calculation of the Speech Intelligibility Index},
  number      = {S3.5-1997 (R2007)},
  year        = {1997},
  institution = {ANSI}
}

@article{ingardising1967,
  author      = {Ingard, Uno and Ising, Hartmut},
  title       = {Acoustic Nonlinearity of an Orifice},
  journal     = {The Journal of the Acoustical Society of America},
  volume      = {42},
  number      = {1},
  pages       = {6--17},
  year        = {1967},
  month       = {Jul},
  issn        = {0001-4966},
  doi         = {10.1121/1.1910576},
  url         = {https://doi.org/10.1121/1.1910576}
}

@article{liu2000locally,
  author      = {Liu, Zhengyou and Zhang, Xixiang and Mao, Yiwei and Zhu, Y. Y. and Yang, Zhiyu and Chan, C. T. and Sheng, Ping},
  title       = {Locally Resonant Sonic Materials},
  journal     = {Science},
  volume      = {289},
  number      = {5485},
  pages       = {1734--1736},
  year        = {2000},
  month       = {Sep},
  issn        = {0036-8075},
  doi         = {10.1126/science.289.5485.1734},
  url         = {https://doi.org/10.1126/science.289.5485.1734}
}

@article{fang2006ultrasonic,
  author      = {Fang, Nicholas and Xi, Dongjuan and Xu, Jianyi and Ambati, Muralidhar and Srituravanich, Werayut and Sun, Cheng and Zhang, Xiang},
  title       = {Ultrasonic metamaterials with negative modulus},
  journal     = {Nature Materials},
  volume      = {5},
  number      = {6},
  pages       = {452--456},
  year        = {2006},
  month       = {Jun},
  issn        = {1476-4660},
  doi         = {10.1038/nmat1644},
  url         = {https://doi.org/10.1038/nmat1644}
}

@article{ma2016acoustic,
  author      = {Ma, Guancong and Sheng, Ping},
  title       = {Acoustic metamaterials: From local resonances to broad horizons},
  journal     = {Science Advances},
  volume      = {2},
  number      = {2},
  pages       = {e1501595},
  year        = {2016},
  month       = {Feb},
  issn        = {2375-2548},
  doi         = {10.1126/sciadv.1501595},
  url         = {https://doi.org/10.1126/sciadv.1501595}
}

@article{cummer2016controlling,
  author      = {Cummer, Steven A. and Christensen, Johan and Al{\`u}, Andrea},
  title       = {Controlling sound with acoustic metamaterials},
  journal     = {Nature Reviews Materials},
  volume      = {1},
  number      = {3},
  pages       = {16001},
  year        = {2016},
  month       = {Feb},
  issn        = {2058-8437},
  doi         = {10.1038/natrevmats.2016.1},
  url         = {https://doi.org/10.1038/natrevmats.2016.1}
}

@article{fano1961effects,
  author      = {Fano, U.},
  title       = {Effects of Configuration Interaction on Intensities and Phase Shifts},
  journal     = {Physical Review},
  volume      = {124},
  number      = {6},
  pages       = {1866--1878},
  year        = {1961},
  month       = {Dec},
  issn        = {0031-899X},
  doi         = {10.1103/physrev.124.1866},
  url         = {https://doi.org/10.1103/physrev.124.1866}
}

@article{qu2025generalized,
  author      = {Qu, Sichao and Yang, Min and Huang, Sibo and Liu, Shuohan and Dong, Erqian and Li, Helios Y. and Sheng, Ping and Abrahams, I. David and Fang, Nicholas X.},
  title       = {Generalized causality constraint based on duality symmetry reveals untapped potential of sound absorption},
  journal     = {Nature Communications},
  volume      = {16},
  number      = {1},
  pages       = {10749},
  year        = {2025},
  month       = {Nov},
  issn        = {2041-1723},
  doi         = {10.1038/s41467-025-65786-w},
  url         = {https://doi.org/10.1038/s41467-025-65786-w}
}

@article{brummund2021occlusion,
  author      = {Carillo, K{\'{e}}vin and Doutres, Olivier and Sgard, Franck},
  title       = {On the removal of the open earcanal high-pass filter effect due to its occlusion: A bone-conduction occlusion effect theory},
  journal     = {Acta Acustica},
  volume      = {5},
  pages       = {36},
  year        = {2021},
  issn        = {2681-4617},
  doi         = {10.1051/aacus/2021029},
  url         = {https://doi.org/10.1051/aacus/2021029}
}

@article{gao2024silencer,
  author      = {Gao, Cong and Hu, Chuandeng and Hou, Bo and Wu, Xiaoxiao and Wen, Weijia},
  title       = {Tunable silencer for rectangular ventilation duct based on composite {Helmholtz} resonators},
  journal     = {Acta Acustica},
  volume      = {8},
  pages       = {22},
  year        = {2024},
  issn        = {2681-4617},
  doi         = {10.1051/aacus/2024013},
  url         = {https://doi.org/10.1051/aacus/2024013}
}

@article{luan2022double,
  author      = {Luan, Yu and Sgard, Franck and N{\'{e}}lisse, Hugues and Doutres, Olivier},
  title       = {A finite element model to predict the double hearing protector effect on an in-house acoustic test fixture},
  journal     = {The Journal of the Acoustical Society of America},
  volume      = {151},
  number      = {3},
  pages       = {1860--1874},
  year        = {2022},
  month       = {Mar},
  issn        = {0001-4966},
  doi         = {10.1121/10.0009835},
  url         = {https://doi.org/10.1121/10.0009835}
}

@article{berger1998development,
  author      = {Berger, Elliott H. and Franks, John R. and Behar, Alberto and Casali, John G. and Dixon-Ernst, Christine and Kieper, Ronald W. and Merry, Carol J. and Mozo, Ben T. and Nixon, Charles W. and Ohlin, Doug and Royster, Julia Doswell and Royster, Larry H.},
  title       = {Development of a new standard laboratory protocol for estimating the field attenuation of hearing protection devices. {Part III}. The validity of using subject-fit data},
  journal     = {The Journal of the Acoustical Society of America},
  volume      = {103},
  number      = {2},
  pages       = {665--672},
  year        = {1998},
  month       = {Feb},
  issn        = {0001-4966},
  doi         = {10.1121/1.423236},
  url         = {https://doi.org/10.1121/1.423236}
}

@article{price1996evaluation,
  author      = {Price, G. Richard and Kalb, Joel T.},
  title       = {Evaluation of hazard from intense sound with a mathematical model of the human ear},
  journal     = {The Journal of the Acoustical Society of America},
  volume      = {100},
  number      = {4, Supplement},
  pages       = {2674},
  year        = {1996},
  month       = {Oct},
  issn        = {0001-4966},
  doi         = {10.1121/1.416935},
  url         = {https://doi.org/10.1121/1.416935},
  note        = {Meeting abstract}
}

@article{price2007validation,
  author      = {Price, G. Richard},
  title       = {Validation of the auditory hazard assessment algorithm for the human with impulse noise data},
  journal     = {The Journal of the Acoustical Society of America},
  volume      = {122},
  number      = {5},
  pages       = {2786--2802},
  year        = {2007},
  month       = {Nov},
  issn        = {0001-4966},
  doi         = {10.1121/1.2785810},
  url         = {https://doi.org/10.1121/1.2785810}
}

@techreport{milstd1474e,
  author      = {{U.S. Department of Defense}},
  title       = {Department of Defense Design Criteria Standard: Noise Limits},
  number      = {MIL-STD-1474E},
  year        = {2015},
  institution = {U.S. Department of Defense}
}

@article{wicher2019gunshot,
  author      = {Skrodzka, Ewa and Wicher, Andrzej and Go{\l}{\k{e}}biewski, Roman},
  title       = {A Review of Gunshot Noise as Factor in Hearing Disorders},
  journal     = {Acta Acustica united with Acustica},
  volume      = {105},
  number      = {6},
  pages       = {904--911},
  year        = {2019},
  month       = {Nov},
  issn        = {1610-1928},
  doi         = {10.3813/aaa.919371},
  url         = {https://doi.org/10.3813/aaa.919371}
}

@article{murphy2012firearm,
  author      = {Murphy, William J. and Flamme, Gregory A. and Meinke, Deanna K. and Sondergaard, Jacob and Finan, Donald S. and Lankford, James E. and Khan, Amir and Vernon, Julia and Stewart, Michael},
  title       = {Measurement of impulse peak insertion loss for four hearing protection devices in field conditions},
  journal     = {International Journal of Audiology},
  volume      = {51},
  number      = {sup1},
  pages       = {S31--S42},
  year        = {2012},
  month       = {Feb},
  issn        = {1499-2027},
  doi         = {10.3109/14992027.2011.630330},
  url         = {https://doi.org/10.3109/14992027.2011.630330}
}

@article{chan2001evaluation,
  author      = {Chan, Philemon C. and Ho, Kevin H. and Kan, Kit K. and Stuhmiller, James H. and Mayorga, Maria A.},
  title       = {Evaluation of impulse noise criteria using human volunteer data},
  journal     = {The Journal of the Acoustical Society of America},
  volume      = {110},
  number      = {4},
  pages       = {1967--1975},
  year        = {2001},
  month       = {Oct},
  issn        = {0001-4966},
  doi         = {10.1121/1.1391243},
  url         = {https://doi.org/10.1121/1.1391243}
}

@article{natarajan2023noise,
  author      = {Natarajan, Nirvikalpa and Batts, Shelley and Stankovic, Konstantina M.},
  title       = {Noise-Induced Hearing Loss},
  journal     = {Journal of Clinical Medicine},
  volume      = {12},
  number      = {6},
  pages       = {2347},
  year        = {2023},
  month       = {Mar},
  issn        = {2077-0383},
  doi         = {10.3390/jcm12062347},
  url         = {https://doi.org/10.3390/jcm12062347}
}

@article{bricault2022silencer,
  author      = {Bricault, Charlie and Meng, Yang and Goud{\'{e}}, S{\'{e}}bastien},
  title       = {Optimization of a silencer design using an helmholtz resonators array in grazing incident waves for broadband noise reduction},
  journal     = {Applied Acoustics},
  volume      = {201},
  pages       = {109090},
  year        = {2022},
  month       = {Dec},
  issn        = {0003-682X},
  doi         = {10.1016/j.apacoust.2022.109090},
  url         = {https://doi.org/10.1016/j.apacoust.2022.109090}
}

@article{huang2019embedded,
  author      = {Huang, Sibo and Fang, Xinsheng and Wang, Xu and Assouar, Badreddine and Cheng, Qian and Li, Yong},
  title       = {Acoustic perfect absorbers via {Helmholtz} resonators with embedded apertures},
  journal     = {The Journal of the Acoustical Society of America},
  volume      = {145},
  number      = {1},
  pages       = {254--262},
  year        = {2019},
  month       = {Jan},
  issn        = {0001-4966},
  doi         = {10.1121/1.5087128},
  url         = {https://doi.org/10.1121/1.5087128}
}

@article{lee2019damped,
  author      = {Lee, Taehwa and Nomura, Tsuyoshi and Iizuka, Hideo},
  title       = {Damped resonance for broadband acoustic absorption in one-port and two-port systems},
  journal     = {Scientific Reports},
  volume      = {9},
  number      = {1},
  pages       = {13077},
  year        = {2019},
  month       = {Sep},
  issn        = {2045-2322},
  doi         = {10.1038/s41598-019-49222-w},
  url         = {https://doi.org/10.1038/s41598-019-49222-w}
}

@article{kim2006hrarray,
  author      = {Kim, SangRyul and Kim, Yang-Hann and Jang, Jae-Hee},
  title       = {A theoretical model to predict the low-frequency sound absorption of a {Helmholtz} resonator array},
  journal     = {The Journal of the Acoustical Society of America},
  volume      = {119},
  number      = {4},
  pages       = {1933--1936},
  year        = {2006},
  month       = {Apr},
  issn        = {0001-4966},
  doi         = {10.1121/1.2177568},
  url         = {https://doi.org/10.1121/1.2177568}
}

@book{kinsler2000fundamentals,
  author      = {Kinsler, L. E. and Frey, A. R. and Coppens, A. B. and Sanders, J. V.},
  title       = {Fundamentals of Acoustics},
  edition     = {4},
  year        = {2000},
  publisher   = {John Wiley \& Sons},
  address     = {New York}
}

@article{moore2016review,
  author      = {Moore, Brian C. J.},
  title       = {A review of the perceptual effects of hearing loss for frequencies above 3 {kHz}},
  journal     = {International Journal of Audiology},
  volume      = {55},
  number      = {12},
  pages       = {707--714},
  year        = {2016},
  month       = {Dec},
  issn        = {1499-2027},
  doi         = {10.1080/14992027.2016.1204565},
  url         = {https://doi.org/10.1080/14992027.2016.1204565}
}

@article{flamme2009estimates,
  author      = {Flamme, Gregory A. and Wong, Adam and Liebe, Kevin and Lynd, James},
  title       = {Estimates of auditory risk from outdoor impulse noise {II}: Civilian firearms},
  journal     = {Noise \& Health},
  volume      = {11},
  number      = {45},
  pages       = {231--242},
  year        = {2009},
  issn        = {1463-1741},
  doi         = {10.4103/1463-1741.56217},
  url         = {https://doi.org/10.4103/1463-1741.56217}
}

@article{calmettes2023multiport,
  author      = {Calmettes, Cyril and Perrey-Debain, Emmanuel and Lefran{\c{c}}ois, Emmanuel and Caillet, Julien},
  title       = {A multi-port scattering matrix formalism for the acoustic prediction in duct networks},
  journal     = {Acta Acustica},
  volume      = {7},
  pages       = {13},
  year        = {2023},
  issn        = {2681-4617},
  doi         = {10.1051/aacus/2023013},
  url         = {https://doi.org/10.1051/aacus/2023013}
}

@article{berchtenbreiter2024mpp,
  author      = {Berchtenbreiter, Benedikt and Renz, Andreas and Becker, Stefan},
  title       = {The potential of additively manufactured porous absorbers in the design of multi-layer microperforated absorbers},
  journal     = {Acta Acustica},
  volume      = {8},
  pages       = {37},
  year        = {2024},
  issn        = {2681-4617},
  doi         = {10.1051/aacus/2024039},
  url         = {https://doi.org/10.1051/aacus/2024039}
}

@article{cavalieri2024pressure,
  author      = {Cavalieri, Th{\'{e}}o and Van Damme, Bart},
  title       = {Sound pressure-dependent acoustic absorption by perforated rigid-frame porous materials},
  journal     = {Acta Acustica},
  volume      = {8},
  pages       = {79},
  year        = {2024},
  issn        = {2681-4617},
  doi         = {10.1051/aacus/2024076},
  url         = {https://doi.org/10.1051/aacus/2024076}
}

@article{zurbrugg2014occlusion,
  author      = {Zurbr{\"u}gg, T. and Stirnemannn, A. and Kuster, M. and Lissek, H.},
  title       = {Investigations on the physical factors influencing the ear canal occlusion effect caused by hearing aids},
  journal     = {Acta Acustica united with Acustica},
  volume      = {100},
  number      = {3},
  pages       = {527--536},
  year        = {2014},
  month       = {May},
  issn        = {1610-1928},
  doi         = {10.3813/AAA.918732},
  url         = {https://doi.org/10.3813/aaa.918732}
}

@article{porschmann2000ownvoice,
  author      = {P{\"o}rschmann, C.},
  title       = {Influences of bone conduction and air conduction on the sound of one's own voice},
  journal     = {Acustica united with Acta Acustica},
  volume      = {86},
  number      = {6},
  pages       = {1038--1045},
  year        = {2000}
}

@article{firestein2023sumrule,
  author      = {Firestein, Chen and Shlivinski, Amir and Hadad, Yakir},
  title       = {Sum rule bounds beyond {Rozanov} criterion in linear and time-invariant thin absorbers},
  journal     = {Phys. Rev. B},
  volume      = {108},
  number      = {1},
  pages       = {014308},
  year        = {2023},
  month       = {Jul},
  publisher   = {American Physical Society},
  doi         = {10.1103/PhysRevB.108.014308},
  url         = {https://link.aps.org/doi/10.1103/PhysRevB.108.014308}
}

@article{bravo2022causally,
  author      = {Bravo, Teresa and Maury, C{\'{e}}dric},
  title       = {Causally-guided acoustic optimization of single-layer rigidly-backed micro-perforated partitions: Theory},
  journal     = {Journal of Sound and Vibration},
  volume      = {520},
  pages       = {116634},
  year        = {2022},
  month       = {Mar},
  issn        = {0022-460X},
  doi         = {10.1016/j.jsv.2021.116634},
  url         = {https://doi.org/10.1016/j.jsv.2021.116634}
}

@article{acher2009fundamental,
  author      = {Acher, O. and Bernard, J. M. L. and Mar{\'{e}}chal, P. and Bardaine, A. and Levassort, F.},
  title       = {Fundamental constraints on the performance of broadband ultrasonic matching structures and absorbers},
  journal     = {The Journal of the Acoustical Society of America},
  volume      = {125},
  number      = {4},
  pages       = {1995--2005},
  year        = {2009},
  month       = {Apr},
  issn        = {0001-4966},
  doi         = {10.1121/1.3081529},
  url         = {https://doi.org/10.1121/1.3081529}
}

\end{document}